\documentclass[aps,notitlepage,superscriptaddress,floatfix,pra,twocolumn,nofootinbib]{revtex4-2}

\usepackage{lipsum}
\usepackage{graphicx}
\usepackage{amsmath,amssymb}
\usepackage{braket}
\usepackage{mathtools}
\usepackage{hyperref}
\hypersetup{
  colorlinks   = true, 
  urlcolor     = blue, 
  linkcolor    = blue, 
  citecolor   = blue 
}
\usepackage{multirow}
\usepackage{color}
\usepackage[normalem]{ulem}
\usepackage{amsfonts}
\usepackage{float}
\usepackage{pdfpages} 
\makeatletter
\usepackage{subfigure}
\usepackage{verbatim}
\usepackage{soul, xcolor}
\usepackage{dsfont}
\usepackage{MnSymbol}

\setstcolor{red}

\def\beq{\begin{equation}}
\def\eeq{\end{equation}}
\def\bea{\begin{eqnarray}}
\def\eea{\end{eqnarray}}

\usepackage{pdfpages} 
\makeatletter
\AtBeginDocument{\let\LS@rot\@undefined}
\makeatother

\begin{document}

\title{Zero noise extrapolation of transport coefficients}

 \author{Hansveer Singh}
\affiliation{Max Planck Institute for the Physics of Complex Systems, 01187 Dresden, Germany}

\author{Pieter W. Claeys}
\affiliation{School of Physics, Trinity College Dublin, Dublin 2, Ireland}

\author{Sarang Gopalakrishnan}
\affiliation{Department of Electrical and Computer Engineering, Princeton University, Princeton, NJ 08544, USA}

\author{Anushya Chandran}
\affiliation{Department of Physics, Boston University, 590 Commonwealth Avenue, Boston, Massachusetts 02215, USA}

\begin{abstract}
NISQ devices routinely estimate properties of the noiseless system from measurements at varying noise strengths $\gamma$. 
These zero-noise extrapolation (ZNE) schemes fail to estimate transport coefficients of conserved quantities when the noise violates the associated conservation laws. 
We show how to extrapolate measurements in noisy, one-dimensional, non-integrable spin chains to obtain the spin transport coefficients governing linear response in the zero-noise limit.
As the wavevector of the Fourier mode of a conserved quantity goes to zero, its decay rate vanishes. The coefficient of the leading momentum dependence defines a transport coefficient in the noiseless system, and an analogous effective coefficient in the noisy one. Naive ZNE implicitly assumes that this effective coefficient is analytic in $\gamma$ as $\gamma\to 0$. Our main result is that universal non-linear corrections to hydrodynamics can produce non-analytic dependence on $\gamma$. In particular, we argue that the effective spin diffusion coefficient of a dissipative Heisenberg chain and the effective subdiffusion coefficient of a dipole-conserving fracton approach their noiseless value as $\gamma^{1/2}$.
We benchmark these analytical predictions numerically.
\end{abstract}

\vspace{1cm}

\maketitle

\section{Introduction}

Zero-noise extrapolation (ZNE) evaluates an observable at a range of noise levels and extrapolates the result to zero noise, in order to estimate the observable's value in the noiseless system. This strategy is used to mitigate hardware errors in noisy quantum devices~\cite{Temme:2017aa,Li:2017aa}, to reduce the computational cost of simulating isolated quantum dynamics~\cite{Rakovszky:2022aa,kuo2024energy,lloyd2024ballistic}, and, to some extent, to probe thermalization in isolated quantum systems~\cite{Mori:2024aa,Prosen_2002,10.21468/SciPostPhys.20.2.061,yoshimura2025}. Despite little theoretical justification, ZNE works well in practice for observables that depend analytically on the noise strength, see e.g.~\cite{kandala2019error,kim2023scalable}.

However, ZNE can fail when studying transport properties~\cite{McCulloch:2026aa,Srivatsa:2024aa}. Transport is—strictly speaking—well defined only in the presence of extensive conserved quantities, and noise generically causes conserved quantities to change in time, so that no transport survives at any non-zero noise strength. Beyond this generic obstruction, ZNE fails for two further, more specific reasons. First, even when noise preserves all conservation laws on average, it can modify the stationary state from the noiseless one, causing ZNE to fail to correctly extrapolate linear response coefficients~\cite{Srivatsa:2024aa}. Second, when noise breaks some (but not all) conservation laws, linear response coefficients can acquire a non-analytic dependence on the noise strength~\cite{McCulloch:2026aa}, a departure from the analytic dependence ZNE assumes. Thus standard transport observables evaluated at non-zero noise may bear no relation to their zero-noise counterparts.

In this manuscript, we develop ZNE for linear response coefficients in the presence of generic noise, and characterize their noise dependence. 
We consider non-integrable, one-dimensional spin chains subject to noise that is modeled by the GKLS master equation~\cite{lindblad1976generators,gorini1976completely} and is of strength $\gamma$. In the zero-noise limit, these chains host diffusive or subdiffusive spin transport at infinite temperature, characterized by a generalized diffusion constant $D_{\mathrm{true}}$. The subdiffusive case arises when the dynamics conserves the total dipole moment, known as fracton dynamics~\cite{FractonHydro,PhysRevLett.125.245303}.
We first show that, for $\gamma\neq0$, one can define effective generalized diffusion constants $D_{\mathrm{eff}}(\gamma)$ through the decay of Fourier modes. For sufficiently small momenta $k$, we argue the decay rate of Fourier modes, $R(k,\gamma)$, takes the form
\begin{equation}
R(k,\gamma) = D_{\mathrm{eff}}(\gamma)k^z+ R(0,\gamma),
\label{eqn:tworate}
\end{equation}
where $z$ is the dynamical exponent and $R(0,\gamma)$ is the decay rate of the total conserved charge. This two-rate ansatz reflects that an effective notion of transport survives even though neither $D_{\mathrm{eff}}(\gamma)$ nor $z$ need converge to their true zero-noise values as $\gamma\to0$. 

If the noise causes the noiseless stationary state to evolve in time, then $D_{\mathrm{eff}}(\gamma)$ should be extracted before the charge density changes appreciably. In numerical simulations, wherein expectation values can be extracted without destroying the time evolved state, later time data can also be put to use. Specifically, we show that in the limit $\gamma\to 0$, $t\to\infty$ holding $\gamma t$ fixed, ZNE recovers the noiseless diffusion constant characterizing linear response about a stationary state with charge density fixed by $\gamma t$.

%

\begin{figure}[!t]
    \centering
    \includegraphics[width=0.4\textwidth]{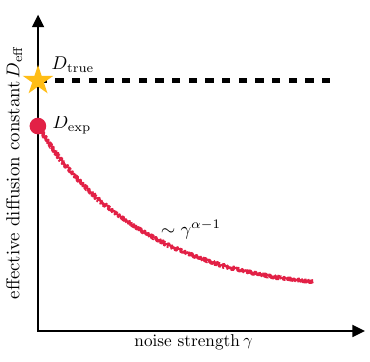}
    \caption{{\bf Zero noise extrapolation (ZNE) of an effective diffusion constant.} The effective diffusion constant (defined in Eq.~\eqref{eqn:tworate}) approaches a zero noise value $D_{\mathrm{exp}}$ as a power law, $\gamma^{\alpha-1}$. When the stationary state depends on noise, $D_{\mathrm{exp}}$ does not equal $D_{\mathrm{true}}$, the true noiseless diffusion constant. Even when $D_{\mathrm{exp}}=D_{\mathrm{true}}$, naive ZNE fails when $\alpha <2$ and $D_{\mathrm{eff}}(\gamma)$ is a non-analytic function of $\gamma$. These non-analytic functions come from universal corrections to hydrodynamics coming from non-linear charge fluctuations. 
     }
    \label{fig:fig1}
\end{figure}

Our main result is that the effective generalized diffusion constant has the following noise dependence (see Fig.~\ref{fig:fig1}),
\begin{equation}
\label{Eq:Deffansatz}
D_{\mathrm{eff}}(\gamma) = D_{\mathrm{exp}} + f_a(\gamma)+f_h(\gamma),
\end{equation}
where $f_a$ is analytic, $f_h$ is non-analytic with a form constrained by the theory of hydrodynamics, and both satisfy $f_a(0) = f_h(0) = 0$ so that $D_{\mathrm{exp}}$ is the zero noise extrapolated value (which equals $D_{\mathrm{true}}$ if the noiseless stationary state remains stationary with noise). We show that $f_h$ is determined by non-linear fluctuations of the conserved quantities at $\gamma=0$, giving a power-law dependence at weak noise, $f_h(\gamma)\sim\gamma^{\alpha-1}$. The exponent $\alpha$ follows from the long-time behavior $\sim t^{-\alpha}$ of current-current correlators and originates from non-linear corrections to hydrodynamics. Standard arguments (e.g. Ref.~\cite{Mukerjee:2006aa}) imply that $\alpha<2$ whenever more than one conserved quantity is present, in which case the non-analytic $f_h$ determines the leading zero-noise behavior of $D_{\mathrm{eff}}$. We illustrate these results by focusing on two models in which analytical and numerical progress is possible. In particular, we show that the effective spin diffusion coefficient of a dissipative Heisenberg chain and the effective subdiffusion coefficient of a dipole-conserving fracton chain approach their noiseless value as $\gamma^{1/2}$.

To simplify the discussion, we first present these arguments for diffusion, before extending these to the subdiffusive case. The rest of the paper is organized as follows: Sec.~\ref{sec:transport_coeffs_intro} reviews the relation between linear response functions and diffusion constants; Sec.~\ref{Sec:transportcoeff} introduces the two-rate ansatz and extracting $D_{\mathrm{eff}}(\gamma)$ from the decay of Fourier modes; Sec.~\ref{Sec:rhoss_rhoeq} applies ZNE when noise modifies the stationary state from the noiseless one; Sec.~\ref{Sec:HydroArg} derives how hydrodynamic contributions appear in $D_{\mathrm{eff}}(\gamma)$; Sec.~\ref{sec:dissipative_Heis} benchmarks our results in a dissipative spin-1/2 Heisenberg chain from~\cite{Claeys:2022aa}; Sec.~\ref{sec:fractons} benchmarks a dissipative spin-1/2 chain hosting fracton dynamics. Conclusions are presented in Sec.~\ref{sec:conc}.

\section{Transport coefficients}
\label{sec:transport_coeffs_intro}
In this section, we review standard linear response relations for the diffusion constant. We assume that the system is one dimensional and we write down formulae for the spin-1/2 case, although they can be straightforwardly generalized. 

This work focuses on computing the spin diffusion constant at infinite temperature and finite chemical potential, i.e. the linear response properties of the equilibrium ensemble, 
\begin{equation}
\rho_{\mathrm{eq}}\propto \exp(-\mu M),
\label{eqn:gc}
\end{equation}
with $M = \sum_x \sigma^{z}_x$ and $\sigma^{z}_x$ the usual Pauli-Z matrix for site $x$. Using the Kubo formula~\cite{doi:10.1143/JPSJ.12.570,RevModPhys.93.025003}, the diffusion constant is given by
\begin{equation}
D = \lim_{t\to\infty}\lim_{L\to \infty}\frac{1}{\chi L}\int_0^{t}d\tau\langle J(\tau)J(0)\rangle,
\end{equation}
where $J(\tau)$ is the time-evolved total current, $\langle A \rangle \equiv \text{tr}(A\rho_{\mathrm{eq}})$, $\chi = \frac{1}{L} (\langle M^2\rangle-\langle M \rangle^2)$ is the normalized static spin susceptibility (the usual static susceptibility is divided by the temperature), and $L$ is the system size. It is useful to also consider the time-dependent diffusion constant, 
\begin{equation}
D(t) = \lim_{L\to \infty}\frac{1}{\chi L}\int_0^{t}d\tau\langle J(\tau)J(0)\rangle.
\label{eqn:kubotime}
\end{equation}
Alternatively, $D(t)$ can be computed using the connected spin-spin structure factor~\cite{RevModPhys.93.025003}, 
\begin{equation}
C_x(t) = \langle \sigma^{z}_x(t)\sigma_0^z(0)\rangle-\langle \sigma^{z}_x\rangle \langle \sigma^z_0\rangle.
\end{equation}
Defining the spatial variance of the charge,
\begin{equation}
\delta x^2(t) \equiv \sum_{x} x^2 C_x(t) - \bigg(\sum_x x C_x(t)\bigg)^2,
\end{equation}
the time-dependent diffusion constant is also given by
\begin{equation}
D(t) = \frac{1}{2}\frac{d}{dt}\delta x^2(t),
\label{eqn:var}
\end{equation}
and consequently, $\frac{d^2}{dt^2}\delta x^2(t) = \lim_{L\to\infty}\frac{2}{\chi L}\langle J(t)J(0)\rangle$. 
Lastly, $D(t)$ can also be computed from the decay rates of Fourier modes. Consider the time-dependent amplitude at wavenumber $k$,
\begin{equation}
C_k(t)\equiv \text{tr}(\sigma^{z}_0\rho_k(t))-\text{tr}( \sigma^{z}_0 \rho_{\mathrm{eq}}), \label{Eq:Cktdef}
\end{equation} 
starting from an initial state $\rho_k(0)$ with a spin density wave at wavenumber $k$, 
\begin{equation}
\rho_k(0)= \left(\mathds{1}+\epsilon \sum_x \cos(kx)\sigma^z_x\right)\rho_{\mathrm{eq}},
\label{eqn:wavestate}
\end{equation}
for $\epsilon\ll 1$. In a diffusive system $C_k(t)\sim e^{-Dk^2t}$, so $D(t)$ can be computed using 
\begin{equation}
D(t) = -\partial_{k^2} \partial_t \log C_k(t)|_{k=0}.
\label{eqn:fourier}
\end{equation}
Evaluating at $k=0$ ensures that any higher order in $k$ corrections to the decay rate vanish. As a final remark, note that $C_k(t)$ is related to $C_x(t)$ via Fourier transform, such that all three definitions are equivalent. Although one can adapt all of these observables to the noisy setting, in the next section we will focus on Eq.~\eqref{eqn:fourier}, before using Eq.~\eqref{eqn:kubotime} to relate the resulting effective diffusion constant to corrections to hydrodynamics.

\section{Effective Transport Coefficients with Noise}
\label{Sec:transportcoeff}
In the presence of noise, the amplitude of spin density waves~\eqref{Eq:Cktdef} still decays, defining an effective diffusion constant. We here motivate the two-rate ansatz for its decay rate, Eq.~\eqref{eqn:tworate}, before extracting the effective diffusion constant from it in later sections.

Consider local and uncorrelated noise, modeled by the GKLS master equation. The density matrix $\rho(t)$ evolves according to
\begin{equation}
\partial_{t}\rho = \mathcal{L}(\rho) = \mathcal{L}_0(\rho)+\gamma \mathcal{D}(\rho),
\label{eqn:lindblad}
\end{equation}
where 
\begin{equation}
\mathcal{D}(\rho) = \sum_{k} L_k\rho L_k^\dagger -\frac{1}{2}\{L_{k}^{\dagger}L_k,\rho\},
\end{equation}
where $L_k$ are Lindblad jump operators and $\mathcal{L}_0$ is the generator of the noiseless system dynamics, e.g. $\mathcal{L}_0(\rho)=-i[H,\rho]$ for Hamiltonian evolution, or $\mathcal{L}_0$ could represent Lindbladian dynamics which preserves the conservation law. For time-independent generators, Eq.~\eqref{eqn:lindblad} implies
\begin{equation}
\rho(t) = e^{\mathcal{L}t}[\rho(0)].
\end{equation}

The noiseless stationary state $\rho_{\mathrm{eq}}$ is defined by $\mathcal{L}_0(\rho_{\mathrm{eq}})=0$. However, since the noisy dissipator $\mathcal{D}$ need not leave this state invariant, we account for this state's time evolution by modifying $C_k(t)$:
\begin{equation}
C_k(t) = \text{tr}(\sigma^{z}_0 \rho_k(t))-\text{tr}(\sigma^z_0 \rho_{\mathrm{eq}}(t)),\label{Eq:Ckmoddef}
\end{equation}
where $\rho_{\mathrm{eq}}(t)=e^{\mathcal{L}t}[\rho_{\mathrm{eq}}]$. Provided the GKLS stationary state is unique (as is the case for all examples here), this modified $C_k(t)$ still decays to zero, but with a rate that depends on both $k$ and $\gamma$,
\begin{equation}
C_k(t) \sim e^{-R(k,\gamma)t}.
\end{equation}
We propose a two-rate ansatz for $R(k,\gamma)$,
\begin{equation}
R(k,\gamma) \sim D_{\mathrm{eff}}(\gamma)k^z+R(0,\gamma),
\label{Eq:repeq1}
\end{equation}
at small $k$, where $z$ is a dynamical exponent ($z=2$ for diffusive dynamics) and $D_{\mathrm{eff}}(\gamma)$ identifies the effective (noisy) diffusion constant. The ansatz is motivated as follows. First, as the noise causes the $k=0$ mode to decay, $R(0,\gamma) > 0$ for $\gamma>0$. Next, as $R(k,0) \sim D_\mathrm{true} k^{z_\mathrm{true}}$ and the noise is short-range correlated, we expect a power-law dependence of $R(k,\gamma)$ on $k$ even at non-zero $\gamma$. In short-range interacting systems, $z=2$ generically, even if $z_\mathrm{true}>2$ and the noiseless dynamics is subdiffusive. This is because generic noise causes incoherent transitions to states forbidden by the subdiffusive dynamics, and allows for an effective diffusion coefficient that vanishes as $\gamma \to 0$.  

The ansatz implies that the effective time-dependent diffusion constant, $D_{\mathrm{eff}}(\gamma,t)$, is given by Eq.~\eqref{eqn:fourier} with the modified $C_k(t)$. $D_{\mathrm{eff}}(\gamma)$ can then be obtained by either: (1) fitting $R(k,\gamma)$ according to Eq.~\eqref{Eq:repeq1} (see e.g. Fig.~\ref{fig:fig2}); or (2) evaluating $D_{\mathrm{eff}}(\gamma,t)$ and extracting its late time value.

We re-iterate that $D_{\mathrm{eff}}(0)$ and $z$ are not required to match their zero noise counterparts, so ZNE of $D_{\mathrm{eff}}(\gamma)$ can fail. Sections~\ref{sec:dissipative_Heis} and~\ref{sec:fractons} provide examples of both cases.

\section{Inequivalence between the noiseless and noisy stationary states}
\label{Sec:rhoss_rhoeq}

The true diffusion constant is a function of the magnetization density, since it is a linear response property of the noiseless equilibrium ensemble, which is itself parametrized by the magnetization density (Eq.~\eqref{eqn:gc}). If the noise causes $\rho_\mathrm{eq}$, and thus the magnetization, to evolve in time, then ZNE becomes subtle as it is not clear which magnetization the extrapolated value corresponds to (see also Ref.~\cite{Srivatsa:2024aa}). We here describe how the right order of limits of $\gamma \to 0$, $t \to \infty$ while holding $\gamma t$ fixed can extrapolate to a predictable value of magnetization density and corresponding diffusion constant $D_{\mathrm{true}}$.

The key insight is that the average total magnetization, $\langle M(t)\rangle$ is only a function of $\gamma t$. Denote $\langle M(t)\rangle$  evaluated at fixed $\gamma t$ as $M^*$ and $D_{\mathrm{eff}}(\gamma,t)$ evaluated at fixed $\gamma t$ as $D_{\mathrm{eff}}(\gamma,M^*)$. Taking $t\to \infty$, $\gamma\to 0$ simultaneously with $\gamma t$ fixed then targets a stationary state, $\rho_{M^*}$,  with an average total magnetization density $M^*/L$. So taking this limit probes the linear response properties of $\rho_{M^*}$, i.e., 
\begin{equation}
\lim_{\substack{t\to\infty\\\gamma\to0\\\gamma t\,\,\text{fixed}}}D_{\mathrm{eff}}(\gamma,M^*) = D_{\mathrm{true}}(M^*),
\label{eqn:diffM}
\end{equation}
where $D_{\mathrm{true}}(M^*)$ is the true diffusion constant for $\rho_{M^*}$ without noise. Note that taking $t\to \infty$ first and then $\gamma \to 0$ without keeping $\gamma t$ fixed targets the diffusion constant obtained by perturbing the steady state with noise $\rho_{\mathrm{ss}}$, i.e. $D_{\mathrm{exp}} = D_{\mathrm{true}}(M_{ss})$ where $M_{ss}$ is the steady state magnetization of the dynamics with noise.

These arguments imply that if $\rho_{\mathrm{eq}}$ is stationary but the initial state, $\rho(0)$, has a different total average magnetization, ZNE can extract the desired zero noise diffusion constants corresponding to different magnetization densities. However, this procedure is not efficient experimentally. Projective measurements collapse the time-evolved state, so that computing expectation values at a time when the system has reached a given magnetization density $M^*$ requires waiting until that (long) time in each run. A more efficient protocol would start from an initial ensemble with average magnetization $M^*$ and evaluate the diffusion constant over a small time window over which the magnetization does not appreciably change. Nevertheless, taking $t\to\infty,\gamma\to0$ simultaneously with $\gamma t$ fixed can still be useful in numerical simulations where expectation values can be evaluated without destroying the state.

\section{Universal Corrections from Hydrodynamics}
\label{Sec:HydroArg}
We now argue that $D_{\mathrm{eff}}(\gamma)$ takes the general form of Eq.~\eqref{eqn:generalform}, reflecting a universal, non-analytic correction from hydrodynamics. The argument rests on two observations. First, generic noise cuts off hydrodynamic processes at the characteristic decay time of the conservation law, $1/\gamma$: only current fluctuations occurring before this cutoff can contribute to $D_{\mathrm{eff}}(\gamma)$. Second, the relevant hydrodynamic processes are controlled by the noisy steady state $\rho_{\mathrm{ss}}$, whose properties depend on the order of limits taken (see Sec.~\ref{Sec:rhoss_rhoeq}). The two observations imply
\begin{equation}
D_{\mathrm{eff}}(\gamma) \approx \lim_{L\to\infty}\frac{1}{\chi_{\mathrm{ss}}L}\int_0^{1/\gamma}d\tau\,\mathrm{tr}(J(\tau)J(0)\rho_{\mathrm{ss}}),
\label{eqn:Deffintegral}
\end{equation}
where $J(\tau)$ is time evolved with the noiseless system dynamics and $\chi_{\mathrm{ss}}$ is the normalized static spin susceptibility of $\rho_{\mathrm{ss}}$. Note that when $\rho_{\mathrm{ss}}=\rho_{\mathrm{eq}}$, Eq.~\eqref{eqn:Deffintegral} reduces to $D_{\mathrm{true}}(t=1/\gamma)$, where $D_{\mathrm{true}}(t)$ is the true time-dependent diffusion constant evaluated in the noiseless stationary state.

Let $\langle J(\tau)J(0)\rangle_{\mathrm{ss}}\equiv \mathrm{tr}(J(\tau)J(0)\rho_{\mathrm{ss}})$. It is useful to rewrite Eq.~\eqref{eqn:Deffintegral} as
\begin{equation}
D_{\mathrm{eff}}(\gamma)\approx D_{\mathrm{exp}} - \frac{1}{\chi_{\mathrm{ss}}L}\int_{1/\gamma}^\infty d\tau\,\langle J(\tau)J(0)\rangle_{\mathrm{ss}},
\label{eqn:Deffsplit}
\end{equation}
where $D_{\mathrm{exp}}=\lim_{L\to\infty}\frac{1}{\chi_{\mathrm{ss}}L}\int_0^{\infty}d\tau\,\mathrm{tr}(J(\tau)J(0)\rho_{\mathrm{ss}})$. The asymptotic behavior of $\langle J(\tau)J(0)\rangle_{\mathrm{ss}}$ dominates the integral in Eq.~\eqref{eqn:Deffsplit}. At late times, $\langle J(\tau)J(0)\rangle_{\mathrm{ss}}\sim \tau^{-\alpha}$. This power law behavior, also known as long-time tails ~\cite{kirkpatrick2002long,dorfman1970velocity,Mukerjee:2006aa,PhysRevA.89.053608,zu2021emergent}, stems from non-linear charge fluctuations that lead to non-linear corrections to hydrodynamics.

To be more precise, let us define coarse-grained (over time/length scales larger than the mean scattering time/mean free path) expectation values of the local charge and current density starting from some initial state, e.g., the perturbed stationary state, as $\mathrm{q}(x,t)$ and $\mathrm{j}(x,t)$, respectively. Fluctuating hydrodynamics postulates that $\mathrm{j}(x,t)$ is related to $\mathrm{q}(x,t)$ via a derivative expansion. In one dimension,
\begin{equation}
\mathrm{j}(x,t) = -D_{\mathrm{exp}}\partial_x \mathrm{q}+\eta(x,t)+\ldots,
\label{eqn:derivexp}
\end{equation}
where $\eta(x,t)$ is noise with zero mean and a variance set by the fluctuation dissipation theorem, i.e $\overline{ \eta(x,t)\eta(x',t')} = 2D_{\mathrm{exp}} \delta(x-x')\delta(t-t')$, where $\overline{A}$ denotes averaging a variable over noise. The higher-order terms that appear are universal, since they are solely determined by symmetries of the dynamics which also leave $\rho_{\mathrm{ss}}$ invariant. However, only terms that are not total derivatives contribute to $\langle J(\tau)J(0)\rangle_{\mathrm{ss}}$, which in the hydrodynamic framework corresponds to $\overline{J(\tau)J(0)}$; therefore only non-linear terms in Eq.~\eqref{eqn:derivexp} contribute (see below for examples). The leading non-linearity determines the exponent $\alpha$, and consequently,
\begin{equation}
D_{\mathrm{eff}}(\gamma)\approx D_{\mathrm{exp}}-\frac{c}{\chi_{\mathrm{ss}}L}\gamma^{\alpha-1},
\label{eqn:powerlaw}
\end{equation}
where $c$ is the coefficient of the leading non-linearity in the derivative expansion. 

Eq.~\eqref{eqn:Deffintegral} is itself only an approximation to $D_{\mathrm{eff}}(\gamma)$, obtained by cutting off the hydrodynamic contribution at $\tau\sim1/\gamma$; it therefore does not capture other, non-universal, analytic-in-$\gamma$ contributions to $D_{\mathrm{eff}}(\gamma)$ that arise from microscopic details of the noise. These analytic contributions dominate whenever $\alpha>2$, since the hydrodynamic correction of Eq.~\eqref{eqn:powerlaw} is then subleading in $\gamma$. This motivates the following general form for $D_{\mathrm{eff}}(\gamma)$,
\begin{equation}
D_{\mathrm{eff}}(\gamma) = D_{\mathrm{exp}}+f_a(\gamma)+f_h(\gamma),
\label{eqn:generalform}
\end{equation}
where $f_a$ is analytic, $f_h$ is determined by Eq.~\eqref{eqn:Deffintegral}, and $f_a(0)=f_h(0)=0$.

The contribution from hydrodynamics determines the leading zero-noise behavior of $D_{\mathrm{eff}}(\gamma)$ (i) when multiple extensive conservation laws, denoted by $Q_i=\sum_x q_x^i$ and coarse grained local charge densities $\mathrm{q}^i(x,t)$, are present, and (ii) the spatial dimension is low enough (here $d=1$). In this case, the leading non-linearity is $\lambda_{ijk}\mathrm{q}^j\partial_x \mathrm{q}^k$, where $\lambda_{ijk}$ is a tensor fixed by the symmetries of the dynamics that leave $\rho_{\mathrm{ss}}$ invariant. Assuming all densities spread diffusively, standard arguments (see Appendix~\ref{app:nonlinearhydro} or e.g. Ref.~\cite{Mukerjee:2006aa}) imply the current-current correlator of the magnetization in one dimension behaves as $\langle J(\tau)J(0)\rangle_{\mathrm{ss}}\sim\tau^{-3/2}$, i.e. $\alpha=3/2$. In higher dimensions, the diffusive tail is weaker ($\alpha = 1+d/2$), so that the non-analytic correction can be subleading to $f_a(\gamma)$.

When only a single conservation law is present, $f_h(\gamma)$ is subleading to $f_a(\gamma)$. Assuming translational invariance and a combined CP symmetry, the lowest-order non-linearity that can contribute to the current-current correlator is $(\partial_x \mathrm{q})^2$. As a result (see again Appendix~\ref{app:nonlinearhydro} or Ref.~\cite{Matthies:2026aa}), $\langle J(\tau)J(0)\rangle_{\mathrm{ss}}\sim\tau^{-5/2}$. With either parity or charge conjugation symmetry, the leading non-linearity takes the form $\mathrm{q}^n(\partial_x \mathrm{q})^m$, with $n$ and $m$ fixed by the symmetries of the dynamics that leave $\rho_{\mathrm{ss}}$ invariant; such terms are higher order than $(\partial_x \mathrm{q})^2$ and can therefore only give a larger $\alpha$.

In the next two sections, we benchmark the predictions of this section and Sec.~\ref{Sec:rhoss_rhoeq} in one-dimensional toy models.

\section{Tests in a Dissipative Heisenberg Chain}
\label{sec:dissipative_Heis}
\begin{figure*}[!t]
    \centering
    \includegraphics[width=\textwidth]{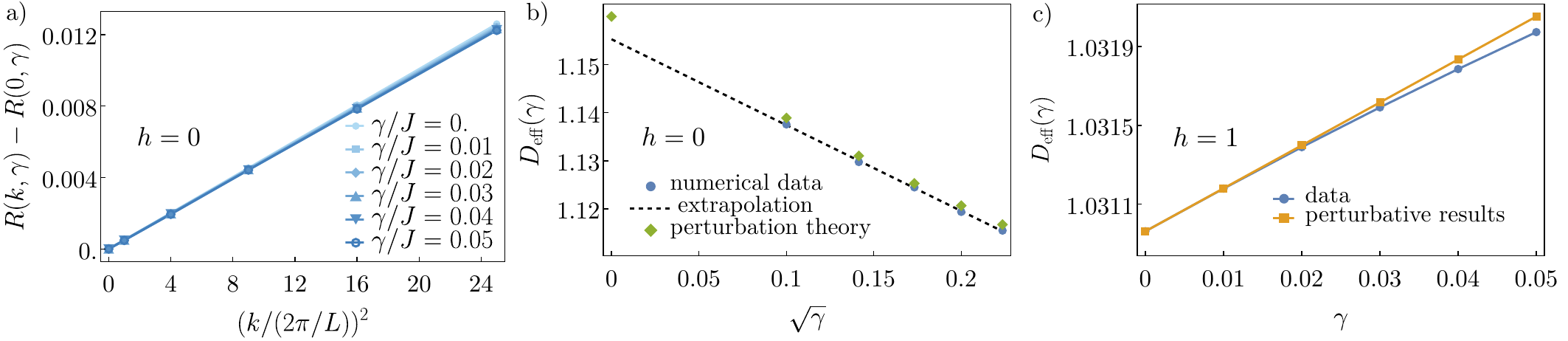}
    \caption{{\bf ZNE for the dissipative Heisenberg chain with depolarization noise without [panels (a) and (b)] and with [panel (c)] a staggered field.} (a) The decay rate $R(k,\gamma)-R(0,\gamma)$ without a staggered field. The largest value of $k$ in this plot is $k= \frac{10\pi}{L} = \frac{\pi}{30}$, i.e. the momenta are quantitatively small. For these small momenta, the rate displays the expected linear dependence on $k^2$ predicted by Eq.~\eqref{Eq:repeq1}. (b) $D_{\mathrm{eff}}(\gamma)$ extracted from $D_{\mathrm{eff}}(\gamma,t)$ using Eq.~\eqref{eqn:fourier}. Saturation occurred around $t/J=30$ for $\gamma=0.03-0.05$ and $t/J=50$ for $\gamma=0.01-0.02$. The data is consistent with the prediction of Eq.~\eqref{eqn:heisdiff} that $D_\mathrm{eff}(\gamma) \sim \sqrt{\gamma}$, and matches the perturbative calculation in $J$ (green diamonds). The extrapolated value, $D_{\mathrm{true}}\approx 1.15$, is consistent with previous estimates and close to the perturbative prediction $D_{\mathrm{true}}\approx 1+J^2=1.16$. (c) $D_{\mathrm{eff}}(\gamma)$ extracted as in panel (b) but in the presence of a staggered field with $h=1$. Saturation of $D_{\mathrm{eff}}(\gamma,t)$ occurs around $t/J=4$. The observed linear dependence at small $\gamma$ is consistent with the prediction for a single conservation law and the perturbative calculation in $\gamma$. In all plots, $J=0.4$ and $L=300$. }
     
    \label{fig:fig2}
\end{figure*}

\begin{figure*}[!t]
    \centering
    \includegraphics[width=.8\textwidth]{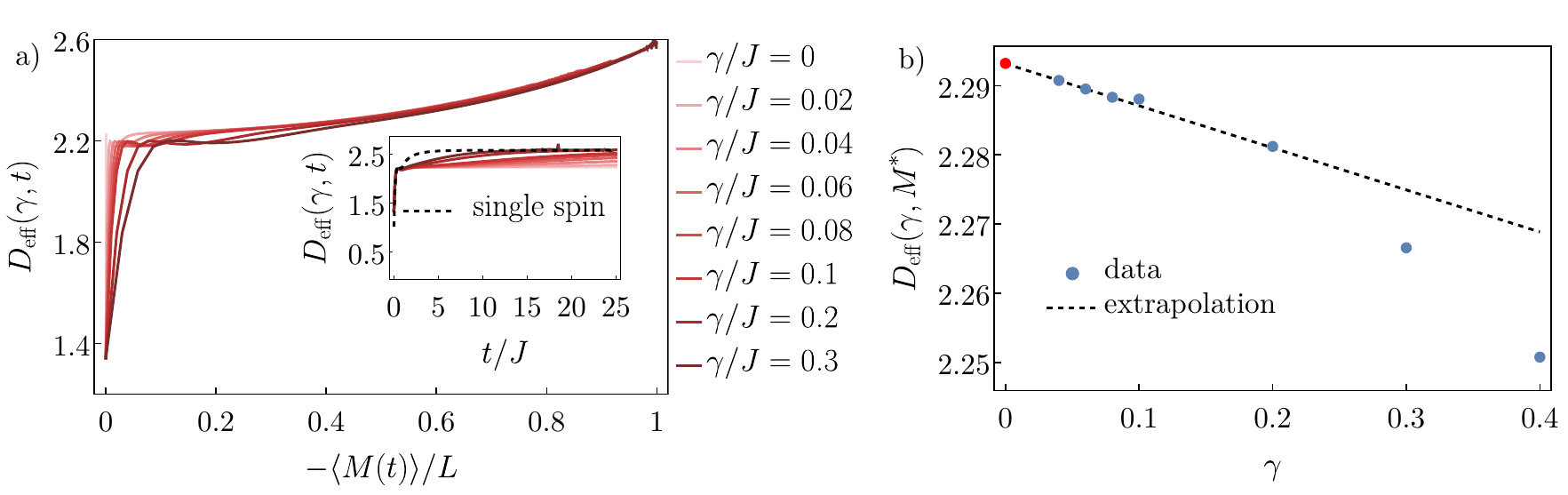}
    \caption{{\bf ZNE for the dissipative Heisenberg chain with amplitude damping noise.} (a) The effective time-dependent diffusion constant, $D_{\mathrm{eff}}(\gamma,t)$, vs. the magnetization density, $\langle M(t)\rangle/L$ for the dissipative Heisenberg model with a staggered field with $h=1$. Regions where the various curves lie on top of each other indicate that $D_{\mathrm{eff}}(\gamma,t)$ has converged to $D_{\mathrm{eff}}(\gamma,M^*/L)$ for a fixed magnetization density $M^*/L$. This occurs for instance when $M^*/L$ approaches $-1$. The inset shows naive ZNE: $D_{\mathrm{eff}}(\gamma,t)$ as $t\to \infty$ approaches its true value at the magnetization density of the steady state, i.e. at $M^*/L=-1$. (b) $D_{\mathrm{eff}}(\gamma,M_*)$ extracted from $D_{\mathrm{eff}}(\gamma,t)$ for various $\gamma$ at a time $t$ for which $M(t)=-L\tanh(1/2)\approx -.4L$.  The extrapolation is only performed for those points where there is approximately linear behavior. The extrapolated value, $D_{\mathrm{true}}(M_*)\approx 2.29$, is consistent with the estimate of the true value from numerical simulations of the zero-noise dynamics for a grand canonical ensemble with $\mu = 1/2$, $D_{\mathrm{true}}(M^*)\approx 2.29$, shown as the red dot. In all plots $J=2$ and $L=300$.} 
    \label{fig:fig3}
\end{figure*}

We illustrate ZNE in the strongly dissipative spin-1/2 Heisenberg chain from~\cite{Claeys:2022aa} using a combination of numerical and perturbative techniques. Figs.~\ref{fig:fig2} and ~\ref{fig:fig3} are our main results. We consider the Heisenberg chain without and with an external staggered field: the former model supports multiple conservation laws, whereas the latter only has a single conservation law.

The dissipative Heisenberg chain consists of a one-dimensional chain of spin-1/2s undergoing SU(2)-symmetric Lindbladian dynamics $\partial_{t}\rho = \mathcal{L}_0(\rho)$, where 
\begin{equation}\label{eq:dissipativeHeisL}
\begin{split}
\mathcal{L}_0(\rho) &= -iJ\sum_{n}[\text{SWAP}_{n,n+1},\rho]\\
&+\sum_{n}(\text{SWAP}_{n,n+1}\rho\,\text{SWAP}_{n,n+1}-\rho).
\end{split}
\end{equation}
Here $\text{SWAP}_{n,n+1}$ is the swap gate for sites $n$ and $n+1$ and $\sigma^{\alpha}_n$ is a Pauli matrix with $\alpha=x,y,z$. The global SU(2) symmetry implies that $\sum_n \sigma^{\alpha}_n$ is conserved for $\alpha=x,y,z$. This Lindbladian models the ensemble dynamics of a Heisenberg chain with exchange couplings fluctuating in space and time around a nonzero mean $J$~\cite{DeNardis:PhysRevLett.127.057201,Claeys:2022aa}. In the limit of a small $J$, this model is amenable to perturbative calculations of the transport properties~\cite{Claeys:2022aa}.

In order to break the SU(2) symmetry and reduce the number of conservation laws to one, we also study the dissipative Heisenberg chain with an additional staggered field:
\begin{equation}
\mathcal{L}_0(\rho) \to \mathcal{L}_0(\rho) -i h\sum_{n}(-1)^n  \left[ \sigma^{z}_n,\rho \right].
\end{equation}

Finally, we subject the chain to either depolarization or amplitude damping noise, each of strength $\gamma$. The respective dissipators are:
\begin{align}
\mathcal{D}_{\mathrm{dep}}(\rho) &= \sum_{n,\alpha}(\sigma^{\alpha}_n\rho \sigma^{\alpha}_n - \rho), \\
\mathcal{D}_{\mathrm{amp}}(\rho) &= \sum_{n} \sigma^{-}_n \rho \sigma^{+}_n - \frac{1}{2}\{\sigma^{+}_n\sigma^{-}_n,\rho\}.
\label{eqn:depeqn}
\end{align}
Above, $\sigma^{\pm}$ are the usual spin raising and lowering operators. Neither type of noise conserves the SU(2) symmetry. The steady state is unique in both cases, corresponding to the uniformly random ensemble with zero average magnetization with depolarization noise, and the maximally polarized state with $M^* = -L$ in the case of amplitude damping. 

The dissipative Heisenberg chain is exactly solvable at $J=0$ for both kinds of noise. In the absence of noise, each Pauli operator $\sigma^{\alpha}_n$ performs an unbiased (continuous-time) random walk, giving rise to diffusive dynamics for the $x,y$ and $z$ magnetization densities with a diffusion constant equal to one in the chosen units (independent of magnetization density)~\cite{Claeys:2022aa}. Thus, $D_\mathrm{true} = 1$ and $z_\mathrm{true}=2$. Noise simply damps the weight of the Pauli operators. In the case of depolarization noise, each $\sigma^{\alpha}_n$ is damped at a rate $4\gamma$ independent of $\alpha$. Consequently, we have that $R(k,\gamma)=D_\mathrm{exp} k^2+4\gamma$, with $D_\mathrm{exp}=1$ matching the zero noise result. For amplitude damping, Pauli operators are damped at different rates and $\sigma^{z}_n$ acquires a finite expectation value in the steady state. Equation~\eqref{Eq:Ckmoddef} subtracts this steady state value. For Fourier modes of $\sigma^z_n$, $R(k,\gamma) = D_\mathrm{exp}k^2+\gamma$ while Fourier modes of $\sigma^{x/y}_n$ have $R(k,\gamma) = D_\mathrm{exp}k^2+\gamma/2$ with $D_\mathrm{exp}=1$.  

For non-zero $J$ and with staggered fields, there are no exact solutions. We therefore turn to perturbation theory and numerical simulations for numerically exact answers. Numerical simulations were performed with ITensor, using tensor network methods~\cite{10.21468/SciPostPhysCodeb.4}. The maximum bond dimension is $\chi=256$ and the truncation error is set to $\epsilon=10^{-16}$, except for the case of the dissipative Heisenberg chain subject to depolarizing noise, where $\epsilon=10^{-23}$.

\subsection{Depolarization Noise}
Consider first the dissipative Heisenberg chain at non-zero $J$ subject to depolarization noise. As discussed above, $R(0, \gamma)=4\gamma$. The uniform equilibrium ensemble (with $M^*=0$) is also the steady state. 
Following Secs.~\ref{Sec:rhoss_rhoeq} and ~\ref{Sec:HydroArg}, we expect that (i) $z=2$ in Eq.~\eqref{eqn:tworate}, (ii) $D_\mathrm{eff}(\gamma)$ extrapolates to the true diffusion constant at $M^*=0$, and (iii) $D_\mathrm{eff}(\gamma)$ depends non-analytically on $\gamma$ as
\begin{equation}
D_{\mathrm{eff}}(\gamma) \approx D_{\mathrm{exp}}-\frac{c}{\chi L}\sqrt{\gamma}.
\label{eqn:heisdiff}
\end{equation}
in the absence of a staggered field, but becomes analytic in $\gamma$ with a staggered field. 

Panel (a) in Fig.~\ref{fig:fig2} confirms that the decay rate of Fourier modes obeys Eq.~\eqref{eqn:tworate} for the dissipative Heisenberg chain subject to depolarization noise with $z=2$. The decay rate $R(k, \gamma)$ is obtained from  $R(k,\gamma) = -\lim_{t\to\infty}\partial_t \log C_k(t)$ with $C_k(t)$ obtained using Eq.~\eqref{Eq:Ckmoddef} where numerically the limit, $t\to\infty$ is obtained by determining when $\partial_t \log C_k(t)$ saturates to a time dependent value. In this case we found good convergence occurring at $t/J=30$ for $\gamma=0.03-0.05$ and $t/J=50$ for $\gamma=0.01-0.02$.

Panel (b) in Fig.~\ref{fig:fig2} plots $D_{\mathrm{eff}}(\gamma)$ versus the predicted leading $\gamma$ dependence, i.e.  $\sqrt{\gamma}$, based on the arguments from Sec.~\ref{Sec:HydroArg}. While the results are consistent with a leading dependence of $\sqrt{\gamma}$, the numerical variation of $D_{\mathrm{eff}}(\gamma)$ is too small to rule out a leading linear dependence on $\gamma$. Thus we supplement our numerical simulations with an analytical perturbative calculation of $D_{\mathrm{eff}}(\gamma)$ which matches Eq.~\eqref{eqn:heisdiff}.

The analytic calculation determines Eq.~\eqref{eqn:heisdiff}  perturbatively in $J$. Working in the Heisenberg picture, $\sigma^{z}_n(t) = e^{\mathcal{L}^{\dagger}t}(\sigma^{z}_n(0))$ evolves according to the adjoint of the Liouvillian. Since Pauli strings, $\mathcal{S}$, form a basis of operators, $\sigma^{z}_n(t) = \sum_{\mathcal{S}} a_{\mathcal{S}}(t)\mathcal{S}$. The equations of motion for $a_{\mathcal{S}}(t)$ are approximated by ignoring contributions from Pauli strings which have a Pauli weight greater than 2, where the Pauli weight is the number of non-identity operators in a Pauli string. Consequently, operators like $\sigma^{z}_n(t)$ are approximated as a sum of Pauli strings of weight 1 and weight 2 at all times. This approach is similar in spirit to dissipation-assisted operator evolution, which has successfully extracted numerical transport coefficients in various settings~\cite{Rakovszky:2022aa,Srivatsa:2024aa,kuo2024energy,lloyd2024ballistic}.
Under this approximation $D_{\mathrm{eff}}(\gamma)$ can be computed analytically as (see Appendix~\ref{app:Perturbation}),
\begin{equation}
D_{\mathrm{eff}}(\gamma) = 1+\frac{J^2}{4}\left(2\gamma+4-\sqrt{2\gamma(2\gamma+8})\right).
\label{eqn:pertexp}
\end{equation}
For $\gamma \ll 1$, this reduces to $D_{\mathrm{eff}}(\gamma) = 1+J^2-J^2\sqrt{\gamma}.$ We recover the form of Eq.~\eqref{eqn:heisdiff}, with $D_{\mathrm{eff}} = 1+J^2$ and $\frac{c}{\chi} = J^2$. 

Panel (b) in Fig.~\ref{fig:fig2} demonstrates that the perturbative calculation and numerical simulations are in good agreement, strengthening the claim of $\sqrt{\gamma}$ being the correct leading order dependence. Additionally, the extrapolated value (obtained by linear in $\sqrt{\gamma}$ extrapolation) of $D_{\mathrm{true}}$ from the numerical simulation is within $\mathcal{O}(J^3)$ from the perturbative calculation.  

Panel (c) in Fig.~\ref{fig:fig2} shows that $D_{\mathrm{eff}}(\gamma)$ is linear in $\gamma$ when a staggered field is present. This is the predicted leading order dependence for a system with a single conservation law, where
\begin{equation}
D_{\mathrm{eff}}(\gamma) \approx D_{\mathrm{exp}}+c \gamma,
\end{equation}
with $c$ a constant. Given that the correction is linear in $\gamma$, perturbation theory in $\gamma$ can serve as an accurate estimate for $D_{\mathrm{eff}}(\gamma)$. Additionally, as $D_{\mathrm{eff}}(\gamma)$ varies only in the third decimal, we supplement numerical results with numerical results for perturbation theory in $\gamma$, rather than in $J$ (see again Appendix~\ref{app:Perturbation}). Panel (c) demonstrates that the numerical results agree with perturbation theory in $\gamma$ for sufficiently small $\gamma/J$, consistent with the linear in $\gamma$ prediction. However, higher order corrections become more noticeable at larger $\gamma/J$, where presumably the corrections from hydrodynamics start to play a role.

\subsection{Amplitude Damping Noise}
For amplitude damping noise the steady state is the fully polarized state, with $M^* = -L$, rather than the infinite temperature ensemble with $M^* = 0$. Naive ZNE fails to extrapolate to $D_{\mathrm{true}}(M^*)$  for $M^*\neq 0 $, but Sec.~\ref{Sec:rhoss_rhoeq} proposes that a different method of extrapolation would succeed. We test this proposal below for the case of the dissipative Heisenberg model with a staggered magnetic field ($h=1$). 

The magnetization evolves in time, when the system is initialized in an equilibrium state of the noiseless system with magnetization $M_\mathrm{eq}$ (Eq.~\eqref{eqn:gc}). Following our discussion in Sec.~\ref{Sec:rhoss_rhoeq}, the magnetization evolves as $\text{tr}(M \rho_{\mathrm{eq}}(t)) = (M_{\mathrm{eq}}-M_{\mathrm{ss}})e^{-\gamma t}+M_{\mathrm{ss}}$.  Consequently, fixed $\gamma t$ corresponds to a fixed magnetization value $M^*$. The expectation is that $D_{\mathrm{eff}}(\gamma,t)$, holding $\gamma t$ fixed (i.e. at fixed $M_*$) while taking $\gamma \to 0 $ and $t \to \infty$, should converge to $D_{\mathrm{true}}(M_*)$. 

Panel (a) of Fig.~\ref{fig:fig3} shows $D_{\mathrm{eff}}(\gamma,t)$ vs $-\langle M(t)\rangle /L$ for various values of $\gamma$. When the curves for different $\gamma$ collapse, $D_{\mathrm{eff}}(\gamma,t)$ is only a function of the magnetization density $M^*/L$.
Consider first magnetization densities $M^*/L\approx -1$ near that of the steady state, when naive ZNE should succeed. The curves in panel (a) lie atop one another in this regime. The late time dynamics of $C_k(t)$ is governed by the single spin flip sector. The inset of panel (a) illustrates naive ZNE: $D_{\mathrm{eff}}(\gamma,t)$, taking $t\to\infty$ before $\gamma\to0$, saturates to the diffusion constant for the ensemble with a single flipped spin obtained from numerical simulations (using exact diagonalization) of the dynamics restricted to the single spin flip sector. Thus ZNE successfully extrapolates the diffusion constant in this case.

Away from this point, Fig.~\ref{fig:fig3}(a) shows several regimes of good convergence. For concreteness, we consider a single point corresponding to $M^* =- L\tanh (\mu)$ with $\mu = 1/2$. Panel (b) illustrates that $D_{\mathrm{eff}}(\gamma,M^*)$ depends linearly on $\gamma$ at small $\gamma$, which eventually breaks down at larger values of $\gamma$. Hence, the extrapolation is only performed for those points which exhibit a linear in $\gamma$ behavior. Because the dynamics is ergodic, $\rho_{M_*}$ should share the same thermodynamic and linear response properties of a grand canonical ensemble, i.e., Eq.~\eqref{eqn:gc} with $\mu = 1/2$. Fig.~\ref{fig:fig3}(b) illustrates that the extrapolation of $D_{\mathrm{eff}}(\gamma,M^*)$ for small $\gamma$ is in good agreement with $D_{\mathrm{true}}(M^*)$ computed from numerically simulating the noiseless dynamics of $\rho_{k}(t)$ with $\rho_{\mathrm{eq}}$ given by Eq.~\eqref{eqn:gc} with $\mu=1/2$ up to a time $t/J=5$. This provides evidence that the ZNE through Eq.~\eqref{eqn:diffM} determines the true diffusion constant as a function of magnetization density.
\section{Fracton Dynamics}
\label{sec:fractons}
\begin{figure*}[!t]
    \centering
    \includegraphics[width=.8\textwidth]{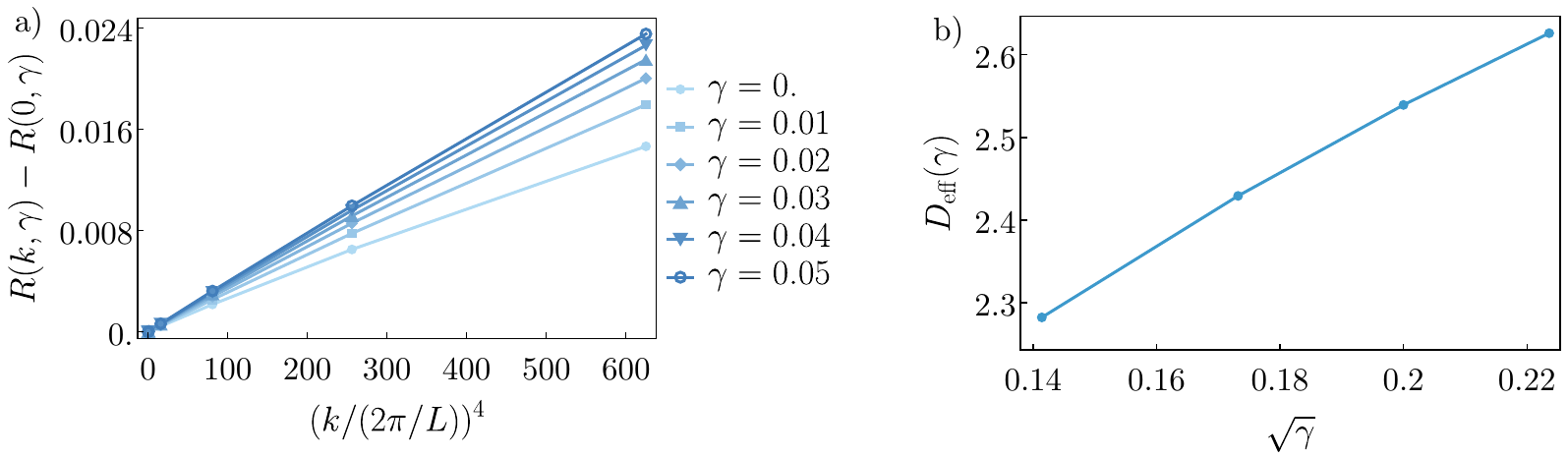}
    \caption{{\bf ZNE for the fracton model with depolarization noise.} (a) The decay rate $R(k,\gamma)-R(0,\gamma)$, where $R(0,\gamma)=4\gamma$. The largest value of $k$ in this plot is $k= \frac{10\pi}{L} = \frac{\pi}{20}$, so that the smallest wavelength is ten times the lattice spacing. For these small momenta, the rate displays the expected linear dependence on $k^4$ predicted by Eq.~\eqref{Eq:repeq1}. (b)  $D_{\mathrm{eff}}(\gamma)$ is approximately a linear function of $\gamma^{1/2}$, consistent with Eq.~\eqref{Eq:subdiffusionconst}. $D_{\mathrm{eff}}(\gamma)$ was extracted from $D_{\mathrm{eff}}(\gamma,t)$ using Eq.~\eqref{eqn:ftsub} and the simulations until $t=29.5$. Fitting the first three points gives an extrapolated value, $D_{\mathrm{true}}\approx 1.55$.
    In all plots $L=100$. Tensor network simulations were performed with a maximum bond dimension $\chi=256$ and with a truncation error of $\epsilon=10^{-23}$.
}
     
    \label{fig:fig4}
\end{figure*}

We illustrate how ZNE generalizes to subdiffusive transport by considering a dissipative spin-1/2 chain that conserves the total dipole moment. The main results of this section are summarized in Fig.~\ref{fig:fig4}.
This figure illustrates the effects of depolarization noise on a chain exhibiting sub-diffusive, so-called fractonic, transport. It provides evidence for Eq.~\eqref{eqn:tworate} with $z=4$ and for Eq.~\eqref{Eq:Deffansatz} with non-analytic corrections in $\gamma$ whose predicted form (derived later in this section) is,  
\begin{equation}
D_{\mathrm{eff}}(\gamma) = D_{\mathrm{exp}}+\frac{c}{\chi L}\gamma^{1/2}.
\label{Eq:subdiffusionconst}
\end{equation}
We restrict ourselves to depolarization noise, which again preserves the noiseless stationary state.

Before moving onto the details of the particular model, we discuss the differences between the diffusive dynamics of the previous section and fractonic dynamics. In one dimension, noiseless fracton dynamics gives rise to sub-diffusive spin transport with a dynamical exponent $z_\mathrm{true}=4$~\cite{FractonHydro,PhysRevLett.125.245303}. The rate of subdiffusion is characterized by a time-dependent subdiffusion constant, $D(t)$, which can be computed using Eq.~\eqref{eqn:kubotime} (see Appendix~\ref{app:fracton}). Additionally  $C_k(t)\sim e^{-Dk^4t}$ so the time-dependent sub-diffusion constant can also be calculated using,
\begin{equation}
D(t) = -\partial_{k^4} \partial_t \log C_k(t)|_{k=0}.
\label{eqn:ftsub}
\end{equation}

An important difference with diffusive dynamics is that the associated continuity equation (written in terms of coarse-grained charge and current variables) becomes
\begin{equation}
\partial_t \mathrm{q} + \partial_x^2 \mathrm{j} = 0
\end{equation}
and the leading terms in the derivative expansion are,
\begin{equation}
\mathrm{j} = D_{\mathrm{true}}\partial_x^2 \mathrm{q}  + \eta(x,t)+ B \mathrm{q}^2\partial_x^2 \mathrm{q} +C'  \mathrm{q} (\partial_x  \mathrm{q})^2+...
\end{equation}
where the noise $\eta(x,t)$ again has zero mean and a variance set by the fluctuation dissipation theorem for linear fracton hydrodynamics $\overline{\eta(x,t)\eta(x',t')}=2D\delta(x-x')\delta(t-t')$.  The latter two terms are the leading higher order terms in the derivative expansion consistent with the symmetries of the system. Importantly, the third term gives rise to $\alpha=3/2$ in Eq.~\eqref{eqn:powerlaw}, and thus to Eq.~\eqref{Eq:subdiffusionconst} (see Appendix~\ref{app:nonlinearhydro}). 

When $\gamma\neq 0$, $D_{\mathrm{eff}}(\gamma,t)$ can be defined using Eq.~\eqref{eqn:ftsub} but, as discussed in Sec.~\ref{Sec:transportcoeff}, the decay rate $R(k,\gamma)$ generically develops a $k^2$ dependence. Depolarization noise is however special in this regard, as it sets the coefficient of the allowed $k^2$ term to be zero. In the Heisenberg picture, depolarization noise only suppresses Pauli strings according to their Pauli weight. It does not generate any strings disallowed by dipole conservation, so that $z=z_\mathrm{true}$. As the noiseless and noisy stationary states coincide at $M^*=0$, $D_\mathrm{exp}=D_\mathrm{true}$. Finally, the ZNE of the subdiffusion constant, using $\alpha=3/2$, leads to Eq.~\eqref{Eq:subdiffusionconst}.

We test the above predictions numerically in a model system. Consider a spin-1/2 chain whose noiseless dynamics is governed by the following Lindbladian,
\begin{equation}
\mathcal{L}_0(\rho) = \sum_{n}(\text{FSWAP}_{n}\rho\,\text{FSWAP}_{n}-\rho ),
\end{equation}
where $\text{FSWAP}_n$ is a five-site operator that causes all transitions that preserve total particle number and total dipole moment with equal amplitude (for brevity, the operator is indexed by the first site). This model has a similar structure as Eq.~\eqref{eq:dissipativeHeisL} in the $J=0$ limit, but constructed to lead to fracton dynamics. Specifically, $\text{FSWAP}_n$ has the following matrix elements,
\begin{equation}
\begin{split}
&\langle \{s'_m\}_{m=n}^{n+4}|\text{FSWAP}_n | \{s_m\}_{m=n}^{n+4}\rangle \\
&\quad= \delta_{\sum_{m=n}^{n+4}s_m,\sum_{m=n}^{n+4}s'_m} \times\delta_{\sum_{m=n}^{n+4}m s_m,\sum_{m=n}^{n+4}ms'_m},
\end{split}
\end{equation}
where $|\{s_m\}_{m=n}^{n+4}\rangle$ denotes a state in the $z$-basis for five consecutive sites, $m=n,n+1,...,n+4$.

The model is, to our knowledge, not exactly solvable. The dynamics is nevertheless simple because $\mathcal{L}_0$ induces a Markov chain in the $z$-basis. The dynamics of the initial state density matrix in Eq.~\eqref{eqn:wavestate} obeys a stochastic master equation,
\begin{equation}
\partial_t |P(t)\rangle = -H|P(t)\rangle,
\end{equation}
where $|P(t)\rangle = \sum_{\{\sigma\}} \rho_{\{\sigma\},\{\sigma\}}(t)|{\{\sigma\}}\rangle$, $\rho_{\{\sigma\},\{\sigma\}}(t)$ is the diagonal element of $\rho$ for the spin configuration $\{\sigma\}\equiv \{\sigma_1,...,\sigma_L\}$ ($\sigma_i=\pm 1$) and 
\begin{equation}
H = \sum_{n}(\mathds{1}-\text{FSWAP}_n) .
\end{equation}

When the strength of the depolarization noise $\gamma\neq0$, this simplification persists. The depolarization noise changes $H$ to
\begin{equation}
H = \sum_{n}(\mathds{1}-\text{FSWAP}_n) +4 \gamma \sum_n |-\rangle\langle -|_n,
\end{equation}
where $|-\rangle_n$ is the -1 eigenstate of $\sigma^{x}_n$.  Despite the evolution being completely classical, as opposed to the models in the previous section, all of our arguments are expected to hold. 

Since the model is not exactly solvable, we use numerical simulations, performed using ITensor, to test the validity of Eq.~\eqref{eqn:tworate}  and Eq.~\eqref{Eq:subdiffusionconst}. 
Panel (a) in Fig.~\ref{fig:fig4} confirms that the decay rate of Fourier modes obeys Eq.~\eqref{eqn:tworate} with $z=4$, confirming the expectation that depolarization noise does not change the dynamical exponent. The decay rate $R(k, \gamma)$ is obtained from  $R(k,\gamma) = -\lim_{t\to\infty}\partial_t \log C_k(t)$ with $C_k(t)$ obtained using Eq.~\eqref{Eq:Ckmoddef}. Panel (b) in Fig.~\ref{fig:fig4} plots $D_{\mathrm{eff}}(\gamma)$ versus the predicted leading $\sqrt{\gamma}$ dependence. While approximately linear, there are deviations at larger values of $\sqrt{\gamma}$. Based on the arguments of Sec.~\ref{Sec:HydroArg}, it is possible that the integrated current-current correlator is cut-off at a time where other higher order nonlinearities are also relevant. For instance, the next leading non-linearity gives an extra $\gamma$ contribution (see Appendix~\ref{app:nonlinearhydro}). For the times and noise strength we can access currently, we cannot ascertain whether such corrections are the correct source of the observed deviations. 

\section{Conclusions}
\label{sec:conc}

We have developed a framework for zero-noise extrapolations (ZNE) of transport coefficients in noisy spin chains, using the decay of Fourier modes to define effective diffusion constants $D_{\mathrm{eff}}(\gamma)$. Naive ZNE of $D_{\mathrm{eff}}(\gamma)$ can fail for two reasons. First, generic noise can drive the system to a stationary state different from the noiseless one, so that the zero-noise and infinite-time limits do not commute. We showed, using numerical simulations and analytical arguments, that taking $\gamma\to0$, $t\to\infty$ with $\gamma t$ fixed instead extracts the true diffusion constant at a magnetization density set by $\gamma t$. While this protocol is not experimentally efficient, it is a useful way to extract the density dependence of the diffusion constant in numerical simulations. Second, even when the stationary state is unchanged, $D_{\mathrm{eff}}(\gamma)$ can depend non-analytically on $\gamma$: universal non-linear corrections to hydrodynamics, arising from non-linear fluctuations of the conserved densities leading to long-time tails decaying $\sim t^{-\alpha}$, produce a $\gamma^{\alpha-1}$ correction that can dominate whenever multiple conservation laws are present ($\alpha<2$). We presented two examples in which such non-analytic corrections arose: a dissipative Heisenberg chain and a dissipative fractonic model. In both cases, naive ZNE succeeds, but its noise dependence encodes universal information about the noiseless dynamics.

It would be interesting to implement our ZNE scheme on data acquired from digital quantum devices. We note that although the ZNE scheme is simple to implement in principle, there are subtleties regarding how one should scale the noise, how one should extrapolate with multiple sources of noise, and what extrapolation method to use in practice without incurring a large cost in sampling overhead~\cite{cai2023quantum}.  

Another interesting follow up direction is to explore transport numerically at finite temperature. A simple protocol would be to prepare the system in a state that is close to the ground state of the target $H$ with an energy density wave of wavenumber $k$ (Eq.~\eqref{eqn:wavestate}), and measure the time-dependent amplitude of the energy density wave during the course of time evolution. The procedure is then repeated for different $k$ and noise strengths $\gamma$. For generic noise, we expect that diffusion coefficients at multiple temperatures are accessible via the procedure in Fig.~\ref{fig:fig3}, even though the stationary state of the noisy dynamics is not a stationary state of the noiseless Hamiltonian with fixed energy density. This procedure to study the equilibrium and linear response properties of finite temperature ensembles is much simpler as compared those Lindblad procedures which prepare exact Gibbs states with complicated multi-body jump operators (see e.g., Ref.~\cite{chen2023quantumthermalstatepreparation, chen2025efficientexactnoncommutativequantum,hahn2025efficient}). 

Lastly, it would be useful to adapt our results to integrable systems. Unlike chaotic systems, corrections to hydrodynamics arise predominantly from properties of the initial state rather than dynamically generated charge fluctuations. Determining the noise dependence in this case remains an open question. 
\section{Acknowledgements}
We thank Curt von Keyserlingk and Giuseppe Del Vecchio Del Vecchio for useful discussions, and Andrew Lucas for providing clarifications regarding the leading non-linearities for fracton dynamics. This work was supported by the Air Force Grant FA9550-24-1-0121 (AC). SG acknowledges support from the National Science Foundation under Award DMR-2516303. AC is grateful to the Max Planck Institute for the Physics of Complex Systems for its hospitality, and AC and PWC acknowledge support from the Max Planck Society. 

\appendix

\begin{widetext}
\section{Deriving long-time tails from non-linear corrections to hydrodynamics}
\label{app:nonlinearhydro}
In this section we give a more detailed explanation of how to obtain long-time tails from non-linear corrections to the derivative expansion in one dimension. In particular we show how to obtain the quoted tails in the main text for the case of multiple conservation laws. The section follows~\cite{Mukerjee:2006aa} closely. 

Focusing on the diffusive case first with multiple conserved charges, denote $Q^a = \sum_x q^a_x$ the conserved charges where the total charge $Q^a$, e.g. the total magnetization in the $x,y,$ and $z$ direction for the dissipative Heisenberg model in the main text. For each conserved charge, $Q^a$ there is an associated total current $J^{a}(t) = \sum_x j^{a}_x(t)$. The coarse-grained charge and current variables are denoted as $\mathrm{q}^a(x,t)$ and $\mathrm{j}^{a}(x,t)$, respectively. We are interested in computing the asymptotic behavior of $\langle J^{a}(t)J^{a}(0)\rangle_{\mathrm{ss}}$ which translates to an average of noise in hydrodynamics, i.e. $\overline{J^{a}(t)J^{a}(0)}$. The leading terms in the derivative expansion are,
\begin{equation}
\mathrm{j}^a = -D \partial_x \mathrm{q^a}+\eta(x,t)+\lambda_{abc}\mathrm{q}^b\partial_x \mathrm{q}^{c}+...
\end{equation}
We can plug this into the expression of the noise averaged current-current correlator to find
\begin{equation}
\overline{J^a(t)J^a(0)} = \int dx dy \lambda_{abc}\lambda_{ade}\overline{ \mathrm{q}^{b}\partial_x \mathrm{q}^{c} \mathrm{q}^{d}\partial_y \mathrm{q}^{e}} = -\lambda_{abc}\lambda_{ade} \int dk_1 dk_3 (k_1 k_3) \overline{ \mathrm{q}^b(k_1,t) \mathrm{q}^c(-k_1,t) \mathrm{q}^d(k_3,0)\mathrm{q}^e(-k_3,0)}. 
\end{equation}
One can now can evaluate the above integral perturbatively by assuming we are at the diffusive fixed point. This means that higher-point correlation functions satisfy Wick's theorem and we have that,
\begin{equation}
\overline{ \mathrm{q}^{a}(k,t)\mathrm{q}^b(k',0)} \sim \delta_{ab}\delta(k+k')e^{-Dk^2t}.
\label{eqn:2pt}
\end{equation}
Making the change of variables $\tilde{k}_i=k_i^2 t$ and noting that the $\delta(k+k')$ contribution will leave only one of the integrals, we see that
\begin{equation}
   \overline{J^a(t)J^a(0)} \sim t^{-3/2}.
\end{equation}


Lastly we discuss the case of subdiffusion in the case of dipole conservation, i.e. so-called fracton hydrodynamics. In fracton hydrodynamics the corresponding current is a two-index tensor, which in one dimension we still denote as $j$, and the corresponding continuity equation reads
\begin{equation}
    \partial_t \mathrm{q} + \partial_x^2 \mathrm{j} = 0.
\end{equation}
Despite this change the subdiffusion constant also obeys the Kubo formula (see Appendix~\ref{app:fracton}). Because of dipole moment conservation, the leading term in the derivative expansion becomes,
\begin{equation}
\mathrm{j} = D\partial_x^2 \mathrm{q} + \eta(x,t) ...
\end{equation}
The dynamics considered in the main text is particle-hole and reflection symmetric. The spin current for fracton hydrodynamics is odd under particle-hole symmetry but even under reflection symmetry. Consequently the leading terms in the derivative expansion are given by,
\begin{equation}
\mathrm{j} = D\partial_x^2 \mathrm{q} + B \mathrm{q}^2 \partial_x^2 \mathrm{q} + C' \mathrm{q} (\partial_x \mathrm{q} )^2+...
\end{equation}
Plugging this into the Kubo formula and going to Fourier space, 
\begin{equation}
\begin{split}
\int dx dy \, \overline{\mathrm{j}(x,t) \mathrm{j}(y,0)} &= \frac{(C')^2}{(2\pi)^4}\int dk_1 dk_2 dk_3 dk_4 \, k_2(k_1+k_2)k_4(k_3+k_4)\\
&\times \overline{ \mathrm{q}(k_1,t)\mathrm{q}(k_2,t)\mathrm{q}(-(k_1+k_2),t)\mathrm{q}(k_3,0)\mathrm{q}(k_4,0)\mathrm{q}(-(k_3+k_4),0)} .
\end{split}
\end{equation}
The form of the propagator changes to
\begin{equation}
\overline{{q}(k,t)\mathrm{q}(k',0)} \sim \delta(k+k')\,e^{-Dk^4t}.
\label{eqn:2pt}
\end{equation}
Using this form of the propagator and following a similar calculation to the previous cases, the leading order decay is
\begin{equation}
\overline{ J(t) J(0)} \rangle\sim t^{-3/2}.
\end{equation}
This leading order decay implies that the leading order $\gamma$ dependence of $D_{\mathrm{eff}}(\gamma)$ is $\sqrt{\gamma}$. In the main text noticeable deviations are seen from this prediction. A possible origin of these deviations is contributions from higher order non-linearities. Consider the leading non-linearity, given by $\mathrm{q}^3 (\partial_x \mathrm{q})^2$. The additional contribution to $\overline{J(t)J(0)}$ from this term is given by
\begin{equation}
\begin{split}
\int dx dy \, \overline{\mathrm{q(x,t)}^3 (\partial_x \mathrm{q(x,t)})^2 \mathrm{q(y,0)}^3 (\partial_y \mathrm{q(y,0)})^2}&= \frac{1}{(2\pi)^{8}}\int \prod_{i=1}^{8} dk_i \, k_4(k_1+k_2+k_3+k_4)k_8(k_5+k_6+k_7+k_8)\\
&\times \langle q(k_1,t)q(k_2,t)q(k_3,t)q(k_4,t)q(-(k_1+k_2+k_3+k_4),t)\\
&\times q(k_5,0)q(k_6,0)q(k_7,0)q(k_8,0)q(-(k_5+k_6+k_7+k_8),0)\rangle.
\end{split}
\end{equation}
Proceeding in the same way as the other cases and using Eq.\eqref{eqn:2pt}, we find $\mathrm{q}^3 (\partial_x \mathrm{q})^2$ modifies the asymptotic behavior to, 
\begin{equation}
\overline{ J(t) J(0)} \rangle \sim At^{-3/2}+Bt^{-2}.
\end{equation}
This scaling implies the $D_{\mathrm{eff}}(\gamma) = D_{\mathrm{true}}+\frac{c}{\chi L}\sqrt{\gamma}+\frac{c'}{\chi L }\gamma+...$.Whether such analytic corrections in $\gamma$ explain the observed deviations is hard to tell with the current data.

\section{Perturbative results for the dissipative Heisenberg chain }
\label{app:Perturbation}
We here derive perturbative expansions in $J$ and $\gamma$ for the dissipative Heisenberg chain.

\subsection{Perturbation theory in $J$}
This section reproduces the $\sqrt{\gamma}$ correction to $D_{\mathrm{eff}}(\gamma)$ at infinite temperature using perturbation theory in $J$. This section closely follows Ref.~\cite{Claeys:2022aa} by performing perturbation theory through a truncation by Pauli weight. For this calculation it is useful to obtain $D_{\mathrm{eff}}(\gamma,t)$ from
\begin{equation}
D_{\mathrm{eff}}(\gamma,t) = -\frac{1}{2}\lim_{k\to0}\partial_k^2\partial_t  C_k(t).
\end{equation}
This agrees with the definition in the main text when $C_k(t)\sim e^{-D k^2t}$.

In the Heisenberg picture, the equation of motion for $C^z_n(t)$, defined as $C^z_n(t)\equiv 2^{-L}\text{tr}(\sigma^z_0(t)\sigma^z_n)$ with $\sigma^z_0(t) = e^{\mathcal{L}^{\dagger}t}[\sigma^z_0(0)]$, is given by
\begin{equation}
\partial_t C^z_n(t) = -2J(C^{xy}_{n,n-1}-C^{xy}_{n-1,n}+C^{xy}_{n,n+1}-C^{xy}_{n+1,n})+\Delta_n C^{z}_n-4\gamma C^{z}_n
\label{eqn:zeom}
\end{equation}
where $\Delta_n C^{z}_n = C^{z}_{n-1}-2C^{z}_n+C^z_{n+1}$ and $C^{xy}_{nm}(t) \equiv 2^{-L}\text{tr}(Z_0(t)\sigma^x_{n}\sigma^y_{m})$. The equations of motion for $C^{xy}_{nm}(t)$ can similarly be determined, which couple both to $C^z_n(t)$ and to Pauli strings with higher weight. However, dealing with the resulting full hierarchy of equations of motion is analytically intractable. Therefore we approximate the equations of motion by---following Ref.~\cite{Claeys:2022aa}---only keeping Pauli strings with weight two or lower.
%
In this approximation, the coupling to Pauli strings with higher weights is discarded and we obtain a closed set of equations,
\begin{equation}
\begin{split}
\partial_tC^{xy}_{nm}(t) &= J(\delta_{n,m+1}(C^{z}_n-C^z_{n+1})+\delta_{n,m-1}(C^{z}_{n-1}-C^{z}_n))+\Delta_{n}C^{xy}_{nm}+\Delta_m C^{xy}_{nm}\\
&+\delta_{|n-m|,1}(C^{xy}_{nm}+C^{xy}_{mn})-8\gamma C^{xy}_{nm}.
\end{split}
\end{equation}
Noting that Eq.~\eqref{eqn:zeom} only depends on $S^{xy}_{nm}\equiv C^{xy}_{nm}-C^{xy}_{mn}$, then
\begin{equation}
\partial_t S^{xy}_{nm}(t) = 2J(\delta_{n,m+1}(C^{z}_n-C^z_{n+1})+\delta_{n,m-1}(C^{z}_{n-1}-C^{z}_n))+\Delta_{n}S^{xy}_{nm}+\Delta_m S^{xy}_{nm}-8\gamma S^{xy}_{nm}.
\label{eqn:seom}
\end{equation}
Fourier transforming Eq.~\eqref{eqn:seom} and~\eqref{eqn:zeom} gives
\begin{align}
\partial_t C^z_k(t) &= -\Omega(k)C^z_k-4\gamma C^z_k -\frac{J}{N}(1-e^{ik})\sum_{q}e^{iq}S^{xy}_{k+q,-q}\,,\\
\partial_t S^{xy}_{kq}(t) &= -(\Omega(k)+\Omega(q))S^{xy}_{kq}-8\gamma S^{xy}_{kq}+4J(\cos(q)-\cos(k))C^{z}_{k+q}\,, \label{eqn:skeom}
\end{align}
where $\Omega(k)=4\sin^2(k/2)$. 

Integrating Eq.~\eqref{eqn:skeom} for $S^{xy}_{kq}e^{(\Omega(k)+\Omega(q)+8\gamma)t}$, plugging the result into the equation of motion for $C^{z}_{k}(t)$ and taking the continuum limit gives
\begin{equation}
\partial_{t} C^{z}_{k}(t) = -\Omega(k)C^{z}_{k}(t)-4\gamma C_k(t)-4J^2\int_{0}^{2\pi}\frac{dq}{2\pi}(\cos(k-q)-\cos(q))^2\int_0^t ds \, e^{(\Omega(k)+\Omega(q)+4\gamma)(s-t)}C^{z}_{q}(s).
\label{eqn:fullzkeom}
\end{equation}
Defining $\tilde{C}_k^z(t)=e^{(\Omega(k)+4\gamma)t}C^z_{k}(t)$, we have that
\begin{equation}
D_{\mathrm{eff}}(\gamma,t) = 1-\frac{1}{2}\partial_t\partial_k^2 \tilde{C}_k(t)|_{k=0}.
\label{eqn:tildeD}
\end{equation}
Using Eq.~\eqref{eqn:tildeD} and~\eqref{eqn:fullzkeom},
\begin{equation}
D_{\mathrm{eff}}(\gamma,t) = 1+4J^2\int \frac{dq}{2\pi} \sin^2(q)\int_0^t ds \, e^{2(\Omega(q)+2\gamma)(s-t)}\tilde{C}^{z}_s(0)
\end{equation}
Using the fact that $\tilde{C}^{z}_{s}(0)=1$ and evaluating the integral over $s$ yields
\begin{equation}
D_{\mathrm{eff}}(\gamma,t) = 1+\frac{J^2}{4}(2\gamma+4-\sqrt{2\gamma(2\gamma+8}))-4J^2 e^{-2\gamma t}\int \frac{dq}{2\pi} \frac{\sin^2(q)}{\Omega(q)+\gamma} e^{-\Omega(q)t}.
\end{equation}
When $t\to\infty$ and $\gamma\ll 1$, we recover the effective diffusion constant
\begin{equation}
D_{\mathrm{eff}}(\gamma)\approx 1+J^2-J^2\sqrt{\gamma}+\mathcal{O}(\gamma^{3/2}).
\end{equation}

\subsection{Perturbation theory in $\gamma$}
This section details the perturbative calculation in $\gamma$, used in the main text to compare against the numerical results for the dissipative Heisenberg chain with a staggered field subject to depolarization noise. The method used is fairly general and is not restricted to the specific model used in the main text. Assume that $\rho_{\mathrm{eq}}$ is the stationary state under the noisy dynamics. The starting point is to assume that $D_{\mathrm{eff}}(\gamma)$ is analytic in $\gamma$. As a result,
\begin{equation}
D_{\mathrm{eff}}(\gamma)\approx D_{\mathrm{eff}}(0)+\gamma \left.\frac{d D_{\mathrm{eff}}}{d\gamma}\right\rvert_{\gamma=0}+\mathcal{O}(\gamma^2).
\end{equation}
Denoting $\frac{d D_{\mathrm{eff}}}{d\gamma}|_{\gamma=0}\equiv D'_{\mathrm{eff}}(0)$, then evaluating the derivative with respect to $\gamma$ returns
\begin{equation}
D'_{\mathrm{eff}}(0) = -\lim_{\gamma\to 0}\lim_{k\to 0}\partial_{t}\partial_{k^2}\frac{\partial_{\gamma}C_k.}{C_k}
\end{equation}
Recall that $C_{k}(t) = \text{tr}(\sigma^z_0\rho_k(t))$ and $\rho_k(t) = e^{(\mathcal{L}_0+\gamma \mathcal{D})t}[\rho_k(0)]$. Using the relation
\begin{equation}
\lim_{\gamma\to 0}\partial_{\gamma}e^{(\mathcal{L}_0+\gamma \mathcal{D})t} = \int_0^t d\tau\, e^{\mathcal{L}_0(t-\tau)}\mathcal{D}\,e^{\mathcal{L}_0\tau},
\end{equation}
gives the following result,  
\begin{equation}
D_{\mathrm{eff}}'(0) = -\lim_{k\to 0}\partial_{k^2}\partial_t \frac{1}{C_k}\int_0^t d\tau\, \text{tr}(\tilde{\rho_{k}}(t;\tau)\sigma^z_0),
\label{eqn:dderiv}
\end{equation}
where $\tilde{\rho_k}(t;\tau) = e^{\mathcal{L}_0(t-\tau)}\mathcal{D}e^{\mathcal{L}_0\tau}[\rho_k(0)]$ and $C_k(t)$ are now evaluated only using $\mathcal{L}_0$, i.e. $\rho_k(t)=e^{\mathcal{L}_0t}[\rho_k(0)]$. We evaluate Eq.~\eqref{eqn:dderiv} numerically for the main text. 
\section{Fracton Linear Response}
\label{app:fracton}
This section shows that the spin subdiffusion constant for fracton hydrodynamics in one dimension satisfies the Kubo formula without noise. Specifically we show the time-dependent spin subdiffusion constant, $D(t)$, for unitary dynamics satisfies,
\begin{equation}
D(t) = \frac{1}{\chi L}\int_0^t d\tau \langle J(\tau)J(0)\rangle 
\end{equation}
with $\langle A \rangle = \text{tr}(A\rho_{\mathrm{eq}})$  (as in the main text) while for purely dissipative dynamics (i.e., only the terms with jump operators are present in the Lindbladian),
\begin{equation}
D(t) = D(0)-\frac{1}{\chi L}\int_0^t d\tau \langle J(\tau)J(0)\rangle,
\end{equation}
where 
\begin{equation}
D(0) = \frac{-1}{24\chi}\sum_x x^4 \text{tr}(\mathcal{L}^{\dagger}(\sigma^{z}_x(0))\sigma^{z}_0).
\end{equation}
Note that we consider general Lindbladian dynamics that conserve total magnetization and total dipole moment. To the authors' knowledge, a general derivation of the Kubo formula for the spin subdiffusion constant for Lindbladian dynamics with dipole conservation could not be found, so the derivation is presented here.

For fracton hydrodynamics in one dimension $C_k(t)\sim e^{-D k^4 t}$, we have that $\partial_k^4 C_k(t)|_{k=0}\sim -24 Dt$, so that,
\begin{equation}
\frac{d}{dt}\delta x^4 = -24D(t) \quad \textrm{with} \quad \delta x^4(t) = \frac{1}{\chi}\sum_{x} x^4 \,\text{tr}(\sigma^{z}_x(t)\sigma^{z}_0\rho_{\mathrm{eq}}).
\end{equation}
To proceed write the Lindbladian which as $\mathcal{L} = \mathcal{L}_H + \mathcal{L}_D$ with $\mathcal{L}_H(\cdot) =-i[H,\cdot]$ and $\mathcal{L}_D(\cdot) \equiv \sum_m L_m\cdot L_m - \frac{1}{2}\{L_m^2,\cdot\}$, where $L_m$ are assumed to be Hermitian. As a reminder, $\mathcal{L}_H$ and $\mathcal{L}_D$ conserve the total magnetization and dipole moment like the dynamics in the main text without noise (e.g. depolarization or amplitude damping noise).
For dipole conserving models~\cite{FractonHydro,PhysRevLett.125.245303},
\begin{equation}
\frac{d}{dt} \sigma^{z}_x(t) = \mathcal{L}^{\dagger}(\sigma^{z}_x(t)) = \mathcal{L}^{\dagger}_H(\sigma^{z}_x(t))+\mathcal{L}_D^{\dagger}(\sigma^{z}_x(t))\equiv -(\Delta j_x^{H}(t)+\Delta j_x^D(t)). 
\label{eqn:adaction}
\end{equation}
where $\Delta_x$ is the discrete lattice Laplacian, i.e. $\Delta f_x = f_{x+1}-2f_x +f_{x-1}$, and $\mathcal{L}^{\dagger}_{H/D}(\sigma^z(t))\equiv -\Delta j_x^{H/D}(t)$. Consider $\frac{d^2}{dt^2} \delta x^4$. This is given by,
\begin{equation}
\frac{d^2}{dt^2} \delta x^4 = \frac{1}{\chi}\sum_{x} x^4 \frac{d^2}{dt^2} \text{tr}(\sigma^{z}_x(t)\sigma^{z}_0(0)\rho_{\mathrm{eq}})= \frac{1}{\chi}\sum_{x} x^4 \text{tr}(\mathcal{L}^{\dagger}(\sigma^{z}_x(t))\mathcal{L}(\sigma^{z}_0(0))\rho_{\mathrm{eq}}),
\end{equation} 
where we used Eq.~\eqref{eqn:adaction}, $\text{tr}(\mathcal{L}^{\dagger}(A)B) = \text{tr}(A\mathcal{L}(B))$. Note that $\mathcal{L}^{\dagger}_H = -\mathcal{L}_H$ while $\mathcal{L}_D^{\dagger} = \mathcal{L}_D$ and $[H,\rho_{\mathrm{eq}}]=  0$ and $[L_m,\rho_{\mathrm{eq}}]=0$. As a result,
\begin{equation}
\frac{d^2}{dt^2} \delta x^4 = \frac{1}{\mathcal{N}}\sum_{x} x^4 \text{tr}((\Delta j^H_x(t)+\Delta j^D_x(t))(-\Delta j^H_0(0)+\Delta j^D_0(0))). 
\label{eqn:dxfourcurrent}
\end{equation}
Focusing only on the terms with $\Delta j^{H}$ in the above, we have
\begin{align}
&-\sum_{x}x^{4}\text{tr}((j^{H}_{x+1}(t)-2j^{H}_x(t)+j^{H}_{x-1}(t))(j^{H}_{1}(0)-2j^{H}_0(0)+j^{H}_{-1}(0)))\\ 
=&-\sum_{x}(6x^4-4(x-1)^4-4(x+1)^4+(x+2)^4+(x-2)^4)\text{tr}( j_x^{H}(t)j_0^H(0))\\
=&-24\sum_x \text{tr}(j^{H}_x(t)j^{H}_0(0)),
\end{align}
where we used the definition of the discrete Laplacian and made use of translation invariance of the dynamics. The calculation is the same for the other terms in Eq.~\eqref{eqn:dxfourcurrent} and one gets,
\begin{equation}
\begin{split}
\frac{d^2}{dt^2} \delta x^4 &= -\frac{24}{\chi}\sum_x (\text{tr}(j_x^{H}(t)j_0^{H}(0))-\text{tr}(j_x^{D}(t)j_0^{D}(0))-\text{tr}(j_x^{H}(t)j_0^{D}(0))+\text{tr}(j_x^{D}(t)j_0^{H}(0)))\\
&= -\frac{24}{\chi}\sum_x (\text{tr}(j_x(t)j^{H}_0(0))-\text{tr}(j_x(t)j^{D}_0(0))),
\end{split}
\end{equation}
where $j_x(t)\equiv j_x^{H}(t)+j_x^{D}(t)$. Using $J(t) = \sum_x j_x(t)$ and making use of translation invariance, we have
\begin{equation}
\frac{d^2}{dt^2} \delta x^4 
= -\frac{24 }{\chi L}(\text{tr}(J(t)J^{H}(0))-\text{tr}(J(t)J^{D}(0))),
\end{equation}
where $J^{H/D}(t) \equiv \sum_x j^{H/D}_x(t)$. Integrating gives,
\begin{equation}
\frac{d}{dt} \delta x^4 
= -24D(0)-\frac{24 }{\chi L}\int_0^t d\tau(\text{tr}(J(\tau)J^{H}(0))-\text{tr}(J(\tau)J^{D}(0))),
\end{equation}
which leads to the result,
\begin{equation}
D(t) = D(0)+\frac{1 }{\chi L}\int_0^t d\tau (\text{tr}(J(\tau)J^{H}(0))-\text{tr}(J(\tau)J^{D}(0))).
\label{eqn:updKubo}
\end{equation}
Note that one can evaluate $D(0)$ from the relation,
$D(0) = -\frac{1}{24}\frac{d}{dt}\delta x^4|_{t=0} = \frac{-1}{24\chi}\sum_x x^4 \text{tr}(\mathcal{L}^{\dagger}(\sigma^{z}_x(0))\sigma^{z}_0)$. We remark that if the Lindbladian was purely Hamiltonian, i.e. $\mathcal{L} = \mathcal{L}_H$ then $\mathcal{L}^{\dagger}_H= -\mathcal{L}_H$ implies $D(0)=0$ but in general $D(0)\neq 0$.

Note that although this was derived under the assumption of fracton hydrodynamics, the calculation holds more generally. However, in the case of the dissipative Heisenberg model discussed in the main text, $j_x^D(t)= \frac{1}{2}(q_x(t)-q_{x-1}(t))$ which implies $J^D(t)$ is a telescoping sum that only receives boundary contributions which are negligible. Thus in the dissipative Heisenberg case $J(t)\approx J^{H}(t)$ and only the first integrated correlator in Eq.~\eqref{eqn:updKubo} remains. Additionally, for the dissipative fracton model explored in the main text only the jump operator terms are present so only the second integrated correlator is present. 

\end{widetext}

\bibliography{refs}
\end{document}